\documentclass[journal,comsoc]{IEEEtran}
\usepackage[T1]{fontenc}
\usepackage{psfrag,amsmath,amssymb,color,cite, amsmath,graphicx}
\usepackage{stfloats}
\usepackage[caption=false,font=footnotesize]{subfig}
\usepackage{mathtools}

\usepackage{algorithm}
\usepackage{mathrsfs}
\usepackage[algo2e,ruled,vlined]{algorithm2e} 
\usepackage{algpseudocode}
\usepackage{pifont}
\usepackage[colorinlistoftodos]{todonotes}

\newtheorem{remark}{Remark}

\newcommand{\beqn}{\begin{equation}}
\newcommand{\eeqn}{\end{equation}}
\newcommand{\beqa}{\begin{eqnarray}}
\newcommand{\eeqa}{\end{eqnarray}}
\newcommand{\beqas}{\begin{eqnarray*}}
\newcommand{\eeqas}{\end{eqnarray*}}
\newcommand{\beqal}{\begin{align}}
\newcommand{\eeqal}{\end{align}}

\presetkeys{todonotes}{color=green!30}{}
\usepackage{array}
\newcolumntype{P}[1]{>{\centering\arraybackslash}p{#1}}
\newcolumntype{M}[1]{>{\centering\arraybackslash}m{#1}}

\begin{document}

\title{Embedded Pilot-Aided Channel Estimation for OTFS in Delay-Doppler Channels}
\author{\IEEEauthorblockN{P. Raviteja, Khoa T. Phan, and Yi Hong\\}
\thanks{P. Raviteja and Yi Hong are with ECSE Department, Monash University, Clayton, VIC 3800, Australia.
Email: \{raviteja.patchava, yi.hong\}@monash.edu. Khoa T. Phan is with CSIT Department, La Trobe University, Bendigo, VIC 3550, Australia. Email: k.phan@latrobe.edu.au.}}
%\IEEEauthorblockA{ECSE Department, Monash University, Clayton, VIC 3800, Australia\\
%Email: \{raviteja.patchava, khoa.phan, yi.hong, emanuele.viterbo\}@monash.edu}}
\maketitle

\begin{abstract}
Orthogonal time frequency space (OTFS) modulation was shown to provide significant error performance advantages over orthogonal frequency division multiplexing (OFDM) in
delay--Doppler channels. In order to detect OTFS modulated data, the channel impulse response needs to be known at the receiver. 
In this paper, we propose embedded pilot-aided channel estimation schemes for OTFS. In each OTFS frame, we arrange pilot, guard, and data symbols in the delay--Doppler plane to suitably avoid interference between pilot and data symbols at the receiver. We develop such symbol arrangements for OTFS over multipath channels with integer and fractional Doppler shifts, respectively. 
At the receiver, channel estimation is performed based on a threshold method and the estimated channel information is used for data detection via a message passing (MP) algorithm. Thanks to our specific embedded symbol arrangements, both channel estimation and data detection are performed within the same OTFS frame with a minimum overhead.
We compare by simulations the error performance of OTFS using the proposed channel estimation and OTFS with ideally known channel information and observe only a marginal performance loss. We also demonstrate that the proposed channel estimation in OTFS significantly outperforms OFDM with known channel information. Finally, we present extensions of the proposed schemes to MIMO and multi-user uplink/downlink. 
\end{abstract}
\section{Introduction} 
%Fifth-generation (5G) mobile systems are expected to accommodate an enormous number of emerging wireless applications with high data rate %requirements (e.g., real-time video streaming, and online gaming, connected and autonomous vehicles etc.). 

Orthogonal frequency division multiplexing (OFDM) is a popular modulation scheme that are currently deployed in 4G long term evolution (LTE) mobile systems. OFDM is known to achieve good robustness and high spectral efficiency for time-invariant frequency selective channels. However, for high-mobility environments such as high-speed railway mobile communications, the channels can be typically time-varying with high Doppler spreads. Under such high Doppler conditions, OFDM is no longer robust and suffers heavy performance degradations. Hence, new modulation schemes that are robust to channel time-variations are being extensively explored. 

Recently, orthogonal time frequency space
(OTFS) modulation was proposed in \cite{Hadani,Hadani1}. OTFS exhibits significant advantages over OFDM in multipath
delay--Doppler channels where each path exhibits a different delay and Doppler shift. 
In particular, the idea of transmission in the delay-Doppler domain was introduced in \cite{Hadani,Hadani1}. The delay--Doppler domain provides as an alternative representation of a time-varying channel geometry due to moving objects (e.g. transmitters, receivers, or reflectors) in the scene. Leveraging on this representation, OTFS multiplexes each information symbol over a two dimensional (2D) orthogonal basis functions, specifically designed to combat the dynamics of time-varying multipath channels. Then the information symbols placed in the delay-Doppler coordinate system can be converted to the standard time-frequency domain used by traditional modulation schemes such as OFDM. More recently, in \cite{Farhang}, a simplified OTFS structure was proposed by including OFDM for time-frequency signal modulation. Its extension to the multiple-input multiple-output (MIMO) case was presented in \cite{reza}.

In general, OTFS uses the delay-Doppler channel response \cite{Hadani,Hadani1,Hadani2} to parameterize the  effects of a time-varying channel on any transmitted waveform. In the delay-Doppler domain, the response captures the dominant scatterers in the channel, with their specific delay and Doppler parameters. In the time-frequency domain, this corresponds to a standard time-varying impulse response. 

Estimating delay-Doppler channel response at the receiver is necessary to perform OTFS detection \cite{Ravi}-\cite{Chock1}. Hence, in \cite{Chock}, \cite{Chock1}, \cite{patent_OTFS}, \cite{Fish}, pilot-aided channel estimation techniques were investigated. 

In \cite{Chock1}, an entire OTFS frame was used for pilot transmission and the estimated channel information was used  for data detection in next  frame. This method may not be effective if the channel estimation becomes outdated in the following frame. In \cite{Chock,Fish}, OTFS channel estimation was conducted in the time--frequency domain, resulting in higher implementation complexity than that of \cite{Chock1,patent_OTFS}, where the channel estimation was conducted in delay--Doppler domain. 
In \cite{patent_OTFS}, channel estimation was considered for OTFS with ideal pulse-shaping waveform over channels with integer Doppler shifts only, i.e., when the channel Doppler taps are aligned to integer delay--Doppler grid.

Motivated by \cite{patent_OTFS}, in this paper, we consider multipath channels with integer and fractional Doppler shifts, respectively\footnote{ Fractional Doppler shifts usually occur with a low Doppler resolution.}.
Under such setting,  we propose an embedded OTFS channel estimation scheme for point-to-point single-input single-output (SISO) system with ideal and rectangular pulse-shaping waveforms, respectively. Specifically, for each OTFS frame, we arrange a single pilot symbol, guard symbols, and data symbols in the delay--Doppler grid to suitably avoid the interferences between pilot and data symbols. At the receiver, channel estimation is performed based on a threshold method and the estimated channel information is used for data detection via a message passing (MP) algorithm in \cite{Ravi}. Depending on the channel and symbol arrangement, the threshold is chosen to optimize the estimation accuracy.
Thanks to our specific embedded symbol arrangements, both channel estimation and data detection are performed within the same OTFS frame with a minimum overhead (1\% for integer Doppler case and 8\% for fractional Doppler case). 

We compare by simulations the performance of OTFS using the proposed channel estimation schemes and OTFS with perfectly known channel information and observe only a marginal performance degradation. Further, we show that OTFS with our channel estimation  significantly outperforms OFDM, with known channel information. 

Finally, we present the extensions of the proposed channel estimation schemes to MIMO and multi-user uplink/downlink. 

The rest of the paper is organized as follows. Section II reviews basic OTFS concepts and results, which lay the foundations for the development of OTFS-based channel estimation schemes in Section III. Numerical results are presented in Section IV. Extensions of the proposed channel estimation schemes to other different OTFS systems are presented in Section V  followed by the conclusions in Section VI. 

\section{OTFS: Basic concepts and results} 
In this section, we first review the basic concepts and results of OTFS from \cite{Hadani,Hadani1}, \cite{Ravi}.   
\subsection{Basic OTFS concepts/notations} 
-- The {\em time--frequency signal plane} is discretized to a $M\times N$ grid (for some integers $N, M >0$)  by sampling time and frequency axes at intervals $T$ (seconds) and $\Delta f$ (Hz), respectively, i.e., 
\beqn 
\Lambda = \bigl\{(nT,m\Delta f),\; n=0,\hdots,N-1, m=0,\hdots,M-1\bigr\} \nonumber 
\eeqn 

%-- A packet burst has duration $NT$ and bandwidth $M\Delta f$. 

-- The modulated {\em time--frequency samples} $X[n,m], n=0,\hdots,N-1, m=0,\hdots,M-1$, are transmitted over an OTFS frame with duration $T_f = NT$ and  bandwidth $B = M\Delta f$. 

-- The delay--Doppler plane is discretized to a $M\times N$ {\em information} grid 
\beqn 
\Gamma = \Biggl\{\left(\frac{k}{NT},\frac{l}{M\Delta f}\right),\; k=0,\hdots,N-1, l=0,\hdots,M-1\Biggr\}, \nonumber 
\eeqn 
where $1/M\Delta f$ and $1/NT$  represent the quantization steps of the delay and Doppler frequency axes, respectively.

\subsection{OTFS mod/demod} 
The modulator first maps a set of $NM$ information symbols $\{x[k,l], k=0,\ldots,N-1, l=0,\ldots, M-1\}$ from a modulation alphabet $\mathbb{A} = \{ a_1, \cdots, a_{Q} \}$ (e.g. QAM symbols) of size $Q$, arranged on the delay--Doppler information grid $\Gamma$, to $X[n,m]$ in the time--frequency domain grid using the {\em inverse symplectic finite Fourier transform} (ISFFT). Next, the {\em Heisenberg transform} is applied to $X[n,m]$ using transmit pulse $g_{\rm tx}(t)$ to create the time-domain signal $s(t)$. 

The signal $s(t)$ is then transmitted over the wireless channel with complex baseband channel impulse response $h(\tau,\nu)$, which characterizes the channel response to an impulse with delay $\tau$ and Doppler $\nu$ \cite{Jakes}.
The received signal $r(t)$ is processed with the {\em Wigner transform} (implementing a receiver filter with an impulse response $g_{\rm rx}(t)$) followed by a sampler, yielding $Y[n,m]$ in the time--frequency domain. We then apply SFFT on $Y[n,m]$  to obtain received symbols $y[k,l]$ in the delay--Doppler domain for symbol detection \cite{Hadani}.  

\subsection{OTFS input--output analysis}
We now look at the relations between received symbols $y[k,l]$ and transmitted symbols $x[k,l]$. 

We assume that $h(\tau,\nu)$ has finite support bounded by $[0,\tau_{\rm max}]$ on the delay axis and $[-\nu_{\rm max},\nu_{\rm max}]$ on the Doppler axis, where $\tau_{\rm max}$ and $\nu_{\rm max}$ are the maximum delay and the maximum Doppler shift among all channel paths. Since  typically  there  are  only  a
small number of reflectors in the channel with  associated delays and Dopplers, very few parameters are needed to model the channel in the delay-Doppler domain. The sparse representation of the channel is
\[
h(\tau,\nu) = \sum_{i=1}^{P} h_i \delta(\tau-\tau_i) \delta(\nu-\nu_i)
\]
where $P$ is the number of propagation paths, $h_i$, $\tau_i$, and $\nu_i$ represent the complex gain, delay, and Doppler shift associated with the $i$-th path, and $\delta(\cdot)$ denotes the Dirac delta function. We denote by  $l_{\tau_i}, k_{\nu_i}$ the delay and Doppler {\em taps} for the $i$-th path (relatively to the delay--Doppler grid $\Gamma$) defined as 
\beqn
\tau_i = \frac{l_{\tau_i}}{M\Delta f},\;\;\nu_i = \frac{k_{\nu_i} + \kappa_{\nu_i}}{NT}
\label{delaytap}
\eeqn 
where $-\frac{1}{2}< \kappa_{\nu_i} \leq \frac{1}{2}$ represents the {\em fractional Doppler}, i.e., the fractional shift from the nearest Doppler tap $k_{\nu_i}$. We do not need to consider fractional delays, since the resolution $1/M\Delta f$ of the time axis is sufficient to approximate the path delays to the nearest sampling points in typical wide-band systems \cite{wc_book}. Let us denote $l_{\tau}$ and $k_{\nu}$ the delay and Doppler taps corresponding to the largest delay $\tau_{\rm max}$ and Doppler $\nu_{\rm max}$. 

We also assume that the pulses $g_{\text{tx}}(t)$ and $g_{\text{rx}}(t)$ are {\em ideal}, meaning that they satisfy the {\em bi-orthogonal property} condition \cite{Hadani}, i.e., the {\em cross-ambiguity function} $A_{g_{\text{rx}}, g_{\text{tx}}}(t,f) = 0$ for $t \in (nT-\tau_{\rm max},nT+\tau_{\rm max})$, $f \in (m\Delta f-\nu_{\rm max},m\Delta f+\nu_{\rm max})$, $\forall n,m$, except for $n=0, m=0$, where $A_{g_{\text{rx}}, g_{\text{tx}}}(t,f) = 1$ with $t \in (-\tau_{\rm max},\tau_{\rm max})$ and $f \in (-\nu_{\rm max}, +\nu_{\rm max})$. The case of non-ideal yet practical rectangular pulses is discussed in Section III. 

%\subsubsection{Delay--Doppler channel sparse representation}

\subsubsection{Integer Doppler shifts} The relation between $y[k,l]$ and  $x[k,l]$ was derived in \cite{Ravi} as  
% \beqa
% &&\!\!\!\!\!\!\!\! \!\!\!\!\!\!\!\! y[k,l]= \sum_{i=1}^{P} h_i e^{-j2\pi\nu_i\tau_i}x[[k - k_{\nu_i}]_N, [l - l_{\tau_i}]_M] + v[k,l] \nonumber \\ 
% \!\!\!\!\!\!\!\!\!\!\!\!\!\!\!\!\!\!\!\!\!\!\!\!\!\!\!\!\!\!&=&\!\!\!\! \sum_{k'=-k_\nu}^{k_\nu}  \sum_{l'=0}^{l_\tau} b[k',l'] \hat h[k',l'] x[[k-k']_N,[l-l']_M] + v[k,l]. \label{integer}
% \eeqa
\begin{align}
y[k,l]=
\!\!\sum_{k'=-k_\nu}^{k_\nu}  \sum_{l'=0}^{l_\tau} b[k',l'] \hat h[k',l']  x[[k-k']_N,&[l-l']_M] \nonumber \\ 
& + v[k,l] \label{integer}
\end{align}
\vspace{-4mm}

\noindent where $\hat h[k',l'] = h[k',l'] e^{-j2\pi \frac{k'}{N T}\frac{l'}{M\Delta f}} $, $b[k',l'] \in \{0,1\}$ is the path indicator, i.e., $b[k',l']=1$ indicates that there is a path with Doppler tap $k'$  and delay tap $l'$ with corresponding path magnitude $\hat h[k',l']$, otherwise, there is no such path, i.e., $b[k',l']=0$ and $\hat h[k',l']=0$. Finally, the term $v[k,l] \sim \mathcal{CN} (0, \sigma^2)$ is an additive white noise with variance $\sigma^2$, and $[\cdot]_N$, $[\cdot]_M$ denote modulo $N$ and $M$ operations, respectively. We have the total number of paths: 
$$
\sum_{k'=-k_\nu}^{k_\nu}  \sum_{l'=0}^{l_\tau} b[k',l']= P. 
$$
%\todo{\tiny Should be $h'_i$ not $a'_i$. please check.}
Each path circularly shifts the transmitted symbols  by the delay and Doppler taps.
\begin{figure*}
\centering
\subfloat[Tx symbol arrangement ($\square$: pilot; $\circ$: guard symbols; $\times$: data symbols)]{%
       \includegraphics[scale=0.81]{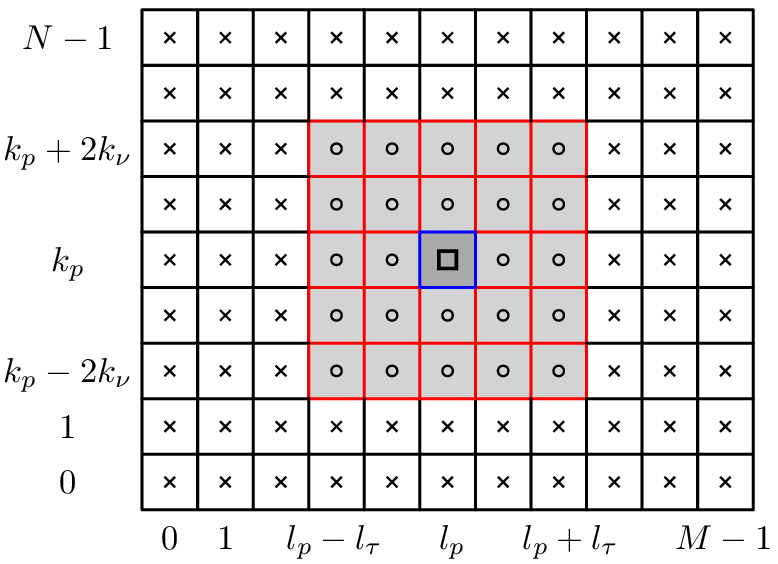}\label{tx_grid}}
    \hspace{7mm}
  \subfloat[Rx symbol pattern ($\triangledown$: data detection, $\boxplus$: channel estimation)]{%
        \includegraphics[scale=0.81]{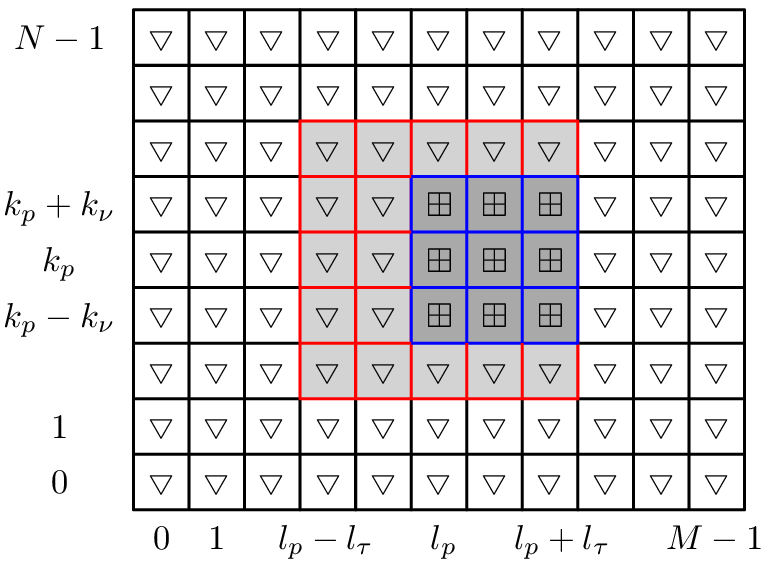}\label{rx_grid}}
\caption{The integer Doppler case}
\end{figure*}
\subsubsection{Fractional Doppler shifts} Similarly,  the following result was derived in \cite{Ravi} for the fractional Doppler case
% \beqn
%  y[k,l]= \sum_{i=1}^{P} \sum_{q=0}^{N-1} \bar h_i(q) x\left[[k-k_{\nu_i}+q]_N, [l - l_{\tau_i}]_M\right] + v[k,l]\label{fractional}
%  \eeqn
\beqa
\!\!\!\!\!\!\!\! \!\!\!\! y[k,l]\!\!\!\!&=&\!\!\!\! \sum_{k'=-k_\nu}^{k_\nu}  \sum_{l'=0}^{l_\tau} b[k',l'] \sum_{q=0}^{N-1} \bar h[k',l',\kappa',q] \nonumber \\ 
&&\quad x\left[[k-k'+q]_N, [l - l']_M\right] + v[k,l]\label{fractional}
\eeqa
where $\kappa'$ denotes the fractional Doppler associated with the $(k',l')$ path, with the path gain 
\[\bar h[k',l',\kappa',q]  = \left(\frac{e^{j {2\pi} (-q - \kappa') }-1}{N e^{j \frac{2\pi}{N} (- q - \kappa')}-N}\right) h[k',l'] e^{-j2\pi\frac{k'+\kappa'}{N T}\frac{l'}{M\Delta f}}.
\]
It can be seen that with fractional Doppler shifts, each received symbol is affected by more  transmitted symbols than in the case of integer Doppler in (\ref{integer}). We can see from (\ref{fractional}) that when $\kappa'=0$, (\ref{fractional}) simplifies to (\ref{integer}) as expected. 
\subsection{OTFS data detection via message passing (MP)} 
From the received symbols $y[k,l]$,  if the channel parameters are known, we can employ the message passing (MP) algorithm in \cite{Ravi} to detect the data symbols $x[k,l]$ using the set of $MN$ linear equations (\ref{integer}) or (\ref{fractional}). %$h_i$, $\tau_i$, and $\nu_i$ (and hence, the corresponding taps $l_{\tau_i}$, $k_{\nu_i}$, and $\kappa_{\nu_i}$) 

%%%%%%%%%%%%%%%%%%%%%%%%%%%%%%%%%%%%%%%%%%%%%%%%%%%%%%%%%%%%%%%%%%%%%%%%%%%
\section{Embedded  channel estimation for point-to-point SISO Case}

We first assume that OTFS with ideal waveforms for multipath channel with integer and fractional Doppler cases. Then we consider the extension to OTFS with practical rectangular waveforms.

\subsection{Integer Doppler Case}
Let $x_p$ denote the pilot symbol with pilot SNR of ${\rm SNR}_{p}$, $x_d[k,l]$ denote the data symbols with data SNR of ${\rm SNR}_{d}$ located at location $[k,l]$ in the delay--Doppler information grid, and $0$ denotes the guard symbol.

%\subsection{Embedded channel estimation}
Motivated by \cite{patent_OTFS}, we place one pilot symbol $x_p$, $N_n$ of the guard symbols, and $MN-N_n-1$ information symbols in the delay--Doppler grid $\Gamma$ for each OTFS frame transmission. The symbols are located in such a way so that at the receiver, we can separate two distinct groups of received symbols: the first group that involves pilot and guard symbols is used for channel estimation, and the second group for data detection. Moreover, the guard symbols guarantee that the received symbols for channel estimation and data detection are not interfered with each other. This helps to provide a more accurate  channel estimation to be used for data detection within the same frame.

%\subsection{Embedded  channel estimation and data detection}
%In this section, we look at the embedded channel estimation for the integer Doppler channel case. 
%\subsection{Transmit symbol arrangement in delay--Doppler grid}
\begin{figure*}
\centering
\subfloat[Tx symbol arrangement ($\square$: pilot; $\circ$: guard symbols; $\times$: data symbols)]{%
       \includegraphics[scale=0.81]{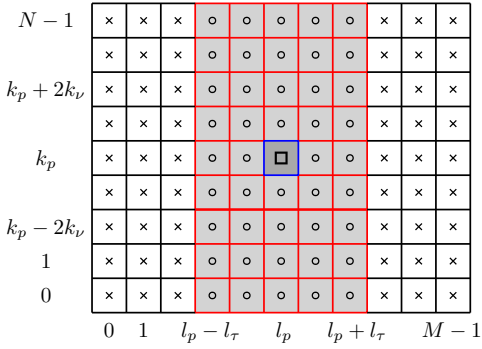}\label{fd_tx_grid_full}}
    \hspace{7mm}
  \subfloat[Rx symbol pattern ($\triangledown$: data detection, $\boxplus$: channel estimation)]{%
        \includegraphics[scale=0.81]{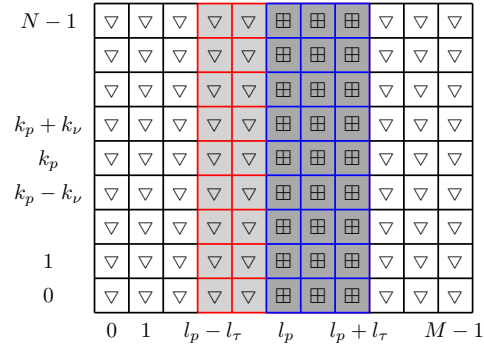}\label{fd_rx_grid_full}}
\caption{The fractional Doppler case: Full guard symbols}
\vspace{-2mm}
\end{figure*}
For a pilot, we first choose arbitrary grid location $[k_p,l_p]$ such that $0 \le k_p \le N-1 $, and $0 \le l_p \le M-1$.  
For ease of representation, we choose $0 \le l_p-l_{\tau}  \le l_p \le  l_p+l_{\tau} \le M-1$, and  $0 \le  k_p-2k_{\nu}  \le k_p \le  k_p+2k_{\nu} \le N-1$. Recall that $l_\tau$ and $k_\nu$ denote the taps corresponding to the maximum delay and Doppler values.

We arrange the pilot, guard, and data symbols in the delay--Doppler grid for an OTFS frame transmission as  in Fig.~\ref{tx_grid}: 
\beqa 
x[k,l] = 
\begin{cases}
x_p &  ~~k=k_p, l =l_p,\\
0 & ~~k_p-2k_{\nu} \le k \le k_p+2k_{\nu},  \\ 
&~~\quad l_p-l_{\tau} \le l \le l_p+l_{\tau},  \\
x_d[k,l] &  ~~\text{otherwise.}
\end{cases} \label{Txinterger}\eeqa 
In this case, we have $N_n=(2l_{\tau}+1)(4k_{\nu}+1)-1$ guard symbols. For example, in Long-Term Evolution (LTE) channels, the overhead for pilot and guard symbols is less than 1\% of the data frame \cite{OTFS_report}.

%\subsection{Received symbols} 
At the receiver, we use the received symbols $y[k,l]$, $k_p - k_{\nu}\le k \le k_p + k_{\nu}, l_p \le l \le l_p+l_{\tau}$ for channel estimation. Then the remaining received symbols $y[k,l]$ on the grid are used for data detection, as shown in Fig. \ref{rx_grid}.   

Due to the transmit symbol arrangement in (\ref{Txinterger}), using (\ref{integer}), we can express the received symbols for channel estimation as
\beqn 
y[k,l] = b[k-k_p,l-l_p] \hat h[k-k_p,l-l_p] x_p + v[k,l]. \label{estimation}
\eeqn 
for $k \in [k_p - k_{\nu}, k_p + k_{\nu}], l \in [l_p, l_p+l_{\tau}]$. 
We can see that if there is a path with  Doppler tap $k-k_p$ and delay tap $l-l_p$, i.e., $b[k-k_p,l-l_p]=1$, we have $y[k,l]  = \hat h[k-k_p,l-l_p] x_p + v[k,l]$. Otherwise, $y[k,l]  =  v[k,l]$. 

Similarly, we can express the received symbols for data detection as in (\ref{integer}), demonstrating no interference between the received symbols for channel estimation and data detection. 

%\subsection{Channel estimation and data detection algorithm} 
We propose a simple channel estimation algorithm as follows. For $k \in [k_p - k_{\nu}, k_p + k_{\nu}], l \in [l_p, l_p+l_{\tau}]$, if the magnitude $|y[k,l]| \ge \mathcal{T}$, where $\mathcal{T}$ is some positive detection threshold, then we estimate $b[k-k_p,l-l_p]=1$ and $\hat h[k-k_p,l-l_p]=y[k,l]/x_p$. Otherwise, we set $b[k-k_p,l-l_p]=\hat h[k-k_p,l-l_p]=0$. 
The proposed threshold-based scheme relies on the fact that if a path exists, the received symbol is the scaled pilot signal with additive white Gaussian noise (see (\ref{estimation})). Otherwise, it is only noise. 

By varying the threshold $\mathcal{T}$, we can alter the miss detection or false alarm probabilities on path detection. As a result, the error performance of data detection is affected by $\mathcal{T}$, as will be shown in Section~\ref{Sec:num_result}. 
 
We then use the estimated information for data detection, i.e., the received symbols $y[k,l]$ for data detection are
\begin{align}
y[k,l] =\!\sum_{k'=-k_\nu}^{k_\nu}  \sum_{l'=0}^{l_\tau} b[k',l'] \hat h[k',l'] x_d[[k-k']_N,&[l-l']_M] \nonumber \\
& + v[k,l] \label{estimate}
\end{align} 
for $k \notin [k_p - k_{\nu}, k_p + k_{\nu}]$ or $ l \notin [l_p,l_p+l_{\tau}]$. Note that we have a total of $MN-(2k_\nu+1)(l_\tau+1)$ received symbols to detect a smaller number of $MN-(2l_{\tau}+1)(4k_{\nu}+1)$ data symbols via the MP algorithm in \cite{Ravi}. 
%%%%%%%%%%%%%%%%%%%%%%%%%%%%%
\subsection{The fractional Doppler case}
%%%%%%%%%%%%%%%%%%%%%%%%%%%%%
We consider two cases using full guard symbols and reduced guard symbols, respectively. The former case offers better channel estimation at the expense of the lower spectral efficiency by using more guard symbols and less data symbols, in contrast to the latter case. 

%%%%%%%%%
\subsubsection{The case with full guard symbols}
%%%%%%%%

%\subsubsection{Transmit symbol arrangement in delay--Doppler grid} 
We arrange the pilot, guard, and data symbols in the delay--Doppler grid, as depicted in Fig. \ref{fd_tx_grid_full}: 
\begin{align}
x[k,l] = 
\begin{cases}
x_p, &  k=k_p, l =l_p\\
0, &  0 \le k \le N-1, l_p \!-\!l_\tau \le l \le  l_p \!+\! l_\tau\\
x_d[k,l], &  \text{otherwise.} \label{Txfrac}
\end{cases} 
\end{align}
For simplicity of notation, we choose $0 \le l_p-l_{\tau}  \le l_p \le  l_p+l_{\tau} \le M-1$. We have the number of guard symbols $N_n=(2l_{\tau}+1)N-1$, and the overhead for pilot and guard symbols is about 8\% in LTE channels  \cite{OTFS_report}.
\begin{figure*}
\centering
\subfloat[Tx symbol arrangement ($\square$: pilot; $\circ$: guard symbols; $\times$: data symbols)]{%
         \includegraphics[scale=0.81]{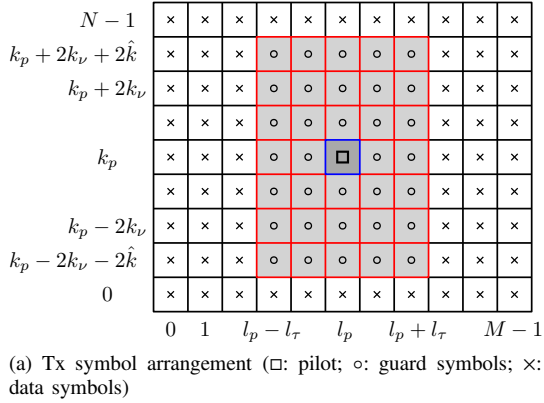}\label{fd_tx_grid_redu}}
    \hspace{7mm}
  \subfloat[Rx symbol pattern ($\triangledown$: data detection, $\boxplus$: channel estimation)]{%
        \includegraphics[scale=0.81]{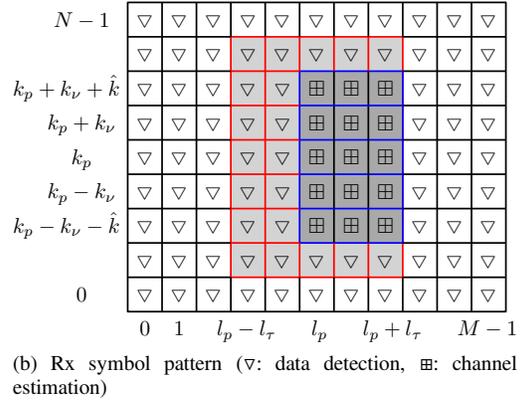}\label{fd_rx_grid_redu}}
\caption{The fractional Doppler case: Reduced guard symbols}
\vspace{-2mm}
\end{figure*}

%\subsubsection{Received symbols} 
At the receiver, we use the received symbols $y[k,l], 0 \le k \le N-1, l_p  \le l \le  l_p + l_\tau$ for channel estimation, and the remaining received symbols $y[k,l]$ for data detection (see Fig. \ref{fd_rx_grid_full}). 

Using (\ref{fractional}), the received symbols $y[k,l]$ for channel estimation are
%W.l.o.g., we set $k_p=0$.  
%From (\ref{fractional}), we can express the received symbols $y[k,l]$ for channel estimation 
\beqn 
y[k,l] \!=\!\!\!\!\!\!
\sum_{k'=-k_\nu}^{k_\nu} \!\!b[k',l\!-\!l_p] \bar h[k',l\!-\!l_p,\kappa',[k_p\!+\!k'\!-\!k]_N\!]
%\right] 
\,x_p %\nonumber \\ 
%&\qquad \qquad \qquad\qquad \qquad \qquad \qquad  
\!\!+\!\! v[k,l] \nonumber 
\eeqn 
for $k \in [0, N-1], l \in [l_p, l_p + l_\tau]$. We can rewrite $y[k,l]$ as 
\beqn 
y[k,l] =  \tilde{b}[l-l_p] \tilde{h}[[k-k_p]_N,l-l_p]x_p + v[k,l] \label{fractional-estimation}
\eeqn 
where  
\begin{align}
\tilde{b}[l-l_p] = 
\begin{cases}
1, &  \sum_{k'=-k_\nu}^{k_\nu} \!\!\!b[k',l-l_p] \ge 1\\
0, &  \text{otherwise}  \nonumber 
\end{cases} 
\end{align}
is the path indicator, and 
\beqn 
\tilde{h}[[k-k_p\!]_N,l-l_p] = \!\!\!\!\sum_{k'=-k_\nu}^{k_\nu} \!\!\!b[k',l-l_p] \bar h[k',l-l_p,\kappa',[k_p\!+\!k'-k]_N\!]\nonumber 
\eeqn 
is the effective path gain from the pilot symbol $x_p$ at location $[k_p,l_p]$ to the received symbol $y[k,l]$.
Then $\tilde{b}[l-l_p]=1$ indicates that there is at least one path with delay tap $l-l_p$, otherwise, $\tilde{b}[l-l_p]=0$. 

%\subsubsection{Channel estimation and data detection algorithm}
Based on (\ref{fractional-estimation}), we propose the following threshold-based channel estimation algorithm. 

For $k \in [0, N-1], l \in [l_p, l_p + l_\tau]$, if $|y[k,l]| \ge \mathcal{T}$, then we have 
$\tilde{b}[l-l_p]=1$, and $\tilde{h}[[k-k_p]_N,l-l_p] = y[k,l]/x_p.$ 
Otherwise, we set 
$ \tilde{b}[l-l_p]=\tilde{h}[[k-k_p]_N,l-l_p] = 0$. Unlike the integer Doppler case, where we estimate whether an individual path with given delay and Doppler taps exists, in this case, we estimate whether there exists {\it at least} one path with a given delay tap. 

For data detection, similar to (\ref{fractional-estimation}), we rewrite (\ref{fractional}) as
% \begin{align}
% & y[k,l]= \sum_{k_{i'}=0}^{N-1}  \sum_{l'=0}^{l_\tau} \sum_{k'=-k_\nu}^{k_\nu} b[k',l']  \bar h(k',l',\kappa',[k'-k_{i'}]_N) \nonumber \\ 
% &\qquad\qquad\qquad\qquad  x\left[[k-k_{i'}]_N, [l - l']_M\right] + v[k,l] \nonumber \\ 
% & = \sum_{k_{i'}=0}^{N-1} \sum_{l'=0}^{l_\tau}\tilde{b}(k_{i'},l')\tilde{h}(k_{i'},l') x([k\!-\!k_{i'}]_N,[l\!-\!l']_M) 
% +v[k,l] \label{estimate1}
% \end{align}
\begin{align}
 y[k,l]= \sum_{l'=0}^{l_\tau}\tilde{b}[l'] \sum_{k'=0}^{N-1} \tilde{h}[k',l'] x_d[[k\!-\!k_{i'}]_N,&[l-l']_M] \nonumber \\
&+v[k,l] \label{estimate1}
\end{align}
for $k \in [0,N-1]$ and $l \notin [l_p, l_p + l_\tau]$. Now we can adapt the MP algorithm in \cite{Ravi} for data detection in (\ref{estimate1}).

Note that, to guarantee no interference between the received symbols for channel estimation and data detection, the guard symbols need to expand over a wider range over the Doppler axis, when compared to the integer Doppler case.

\subsubsection{The case of reduced guard symbols}  
Employing full guard symbols to avoid interferences provide more accurate channel estimation but with reduced spectral efficiency. To improve the spectral efficiency, we can reduce the number of guard symbols and thus increase the number of data symbols, as discussed below.

%\subsubsection{Transmit symbol arrangement in delay--Doppler grid}
We arrange the symbols as in Fig. \ref{fd_tx_grid_redu}
\beqas
x[k,l] = 
\begin{cases}
x_p &  ~~k=k_p, l =l_p,\\
0 & ~~k_p-2k_{\nu}-2 \hat k \le k \le k_p+2k_{\nu} + 2 \hat k,  \\ 
&~~\quad l_p-l_{\tau} \le l \le l_p+l_{\tau},  \\
x_d[k,l] &  ~~\text{otherwise}
\end{cases} \eeqas 
for some integer $\hat k$. For smaller $\hat k$, less guard and more data symbols are used, resulting in an increased spectral efficiency. 

%\subsubsection{Received symbols} 
The received symbols $y[k,l], k_p-k_{\nu}- \hat k \le k \le k_p+k_{\nu} +  \hat k, l_p  \le l \le  l_p + l_\tau$ are used for channel estimation, while the remaining $y[k,l]$ are used for data detection (see Fig. \ref{fd_rx_grid_redu}) 

From (\ref{fractional}), for channel estimation, we have 
% \begin{align}
% \!\! \! y[k,l] & \!=\! \left[\sum_{k'=-k_\nu}^{k_\nu} \!\! b[k',l-l_p] \bar h(k',l-l_p,\kappa',k'-k)\right] x_p \!+\! v[k,l] \nonumber \\
% & =  \tilde{b}(k,l-l_p) \tilde{h}(k,l-l_p)x_p + v[k,l]. \label{fractional-estimation}
% \end{align}
\begin{align}
y[k,l] = %\left[
\tilde{b}[l\!-\!l_p] \tilde{h}[[k\!-\!k_p]_N,l\!-\!l_p]x_p \!+\! \mathcal{I}[k,l] \!+\! v[k,l] \label{fractional-estimation-1}
\end{align}
for $k_p-k_{\nu}- \hat k \le k \le k_p+k_{\nu} +  \hat k, l_p  \le l \le  l_p + l_\tau$. The second term $\mathcal{I}[k,l]$ is the interferences from all neighboring data symbols $x_d[k,l]$, i.e.,
\beqa
\mathcal{I}[k,l]&=&\!\!\!\! \sum_{k'=-k_\nu}^{k_\nu}  \sum_{l'=0}^{l_\tau} b[k',l']\hspace{-7mm} \sum_{q \notin [k_p-2k_{\nu}- 2\hat k, k_p+2k_{\nu} + 2\hat k]} \hspace{-7mm} \bar h[k',l',\kappa',q] \nonumber \\ 
&&\quad x_d\left[[k-k'+q]_N, [l - l']_M\right] 
\eeqa
We observe that the interference $\mathcal{I}[k,l]$ gets larger for smaller $\hat k$, and similarly for the interference from pilot symbols to the received symbols for data detection.  

%\subsubsection{Channel estimation and data detection algorithms}
Similar to the case of full guard symbols, we develop a threshold-based algorithm to estimate $\tilde{b}[l-l_p]$ and $ \tilde{h}[[k-k_p]_N,l-l_p]$ based on (\ref{fractional-estimation-1}) by treating $\mathcal{I}[k,l]$ as additive noise. %We remark that we take the inferences into consideration by employing a larger threshold than that of the full guard symbols case, otherwise the channel estimation and data detection may be less accurate. Nevertheless, 
Based on the simulation results (see next section), we demonstrate that the performance gap of the full guard symbols case (8\% overhead) and reduced guard symbols case (2\% overhead) is indeed marginal. 

\subsection{OTFS with rectangular waveforms} 
So far, we have assumed ideal transmit $g_{\text{tx}}(t)$  and receive $g_{\text{rx}}(t)$  pulses. Since the ideal pulses cannot be realized in practice, we now investigate OTFS with the more practical rectangular pulses at both transmitter and receiver. Although these pulses do not satisfy the bi-orthogonality conditions \cite{Ravi1}, we show that the proposed embedded channel estimation schemes can also be employed for this case. 

Consider the integer Doppler case for simplicity. With rectangular pulses, the input-output symbol relationship in \cite{Ravi1} can be rewritten as  
\begin{align}
y[k,l]=
\!\!\!\sum_{k'=-k_\nu}^{k_\nu}  \sum_{l'=0}^{l_\tau} b[k',l'] \hat h[k',l']\alpha[k,l]  x[[k\!-\!k']_N,&[l\!-\!l']_M\!] \nonumber \\
& + v[k,l] \nonumber 
\end{align}
where
\begin{align*}
\alpha[k,l] & = 
\begin{cases}
e^{j2\pi \left(\frac{l-l'}{M}\right)  \frac{k'}{N} } &  l' \leq l < M\\
\frac{N-1}{N} e^{j2\pi \left(\frac{l-l'}{M}\right)  \frac{k'}{N} } e^{-j 2 \pi \left(\frac{[k-k']_N}{N} \right)} &  0\leq l<l'.
\end{cases}
\end{align*}
Hence, the threshold-based channel estimation technique can be straightforwardly employed by introducing a known phase $\alpha[k,l]$ in the detection process. The thresholds for the rectangular waveforms remains the same as the ideal waveforms, since the channel differs only by a phase.

%%%%%%%%%%%%%%%%%%%%%%%%%%
\section{Numerical results}\label{Sec:num_result}
%%%%%%%%%%%%%%%%%%%%%%%%%
We illustrate the performance in term of bit-error-rate (BER)  of the uncoded OTFS using the proposed channel estimation schemes for integer and fractional Doppler cases. 
We adopt the following system parameters: Carrier frequency of $4$ GHz, sub-carrier spacing of $15$ KHz, $M$ = $512$, $N = 128$, and $4-$QAM signaling.   
For both OTFS and OFDM systems, Extended Vehicular A model \cite{LTE} is used, and each delay tap has a single Doppler shift generated by using Jakes' formula, i.e., $\nu_i = \nu_{\rm{max}}\cos(\theta_i)$, where $\nu_{\rm max}$ is the maximum Doppler shift determined by the UE speed and $\theta_i$ is uniformly distributed over $[-\pi,\pi]$.
\subsection{The integer Doppler case}
%We first consider channel estimation for the integer Doppler case.

\begin{figure}
\centering
\includegraphics[width=2.8in,height=2in]{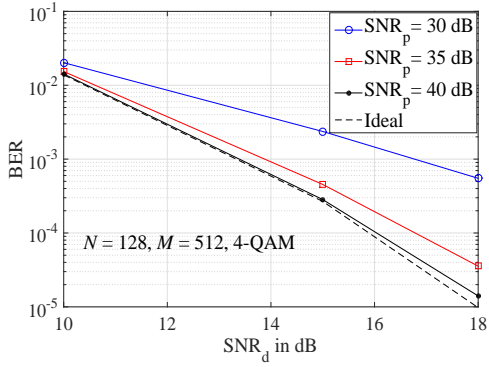}
\caption{BER versus ${\rm SNR}_d$: Integer Doppler case.}
\label{sim1}
\end{figure}

\begin{figure}
\centering
\includegraphics[width=2.8in,height=2in]{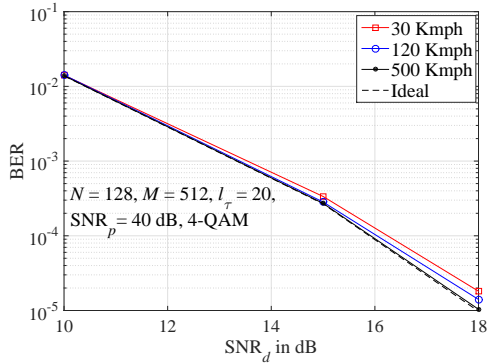}
\caption{BER versus ${\rm SNR}_d$ for different Dopplers}
\label{sim4}
\end{figure}

\begin{figure}
\centering
\includegraphics[width=2.8in,height=2in]{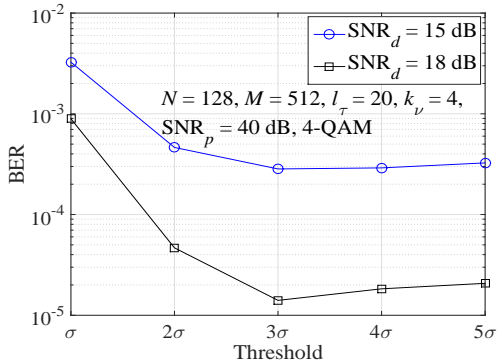}
\caption{BER versus channel estimation thresholds: Integer Doppler case.}
\label{sim2}
\end{figure}  

Fig. \ref{sim1} compares BER versus data SNRs (${\rm SNR}_d$) for OTFS with known channel information (ideal case) and OTFS using the proposed channel estimation for the integer Doppler case with ${\rm SNR}_p= 30,$ $35,$ and $40$ dB and $\mathcal{T}=3\sigma$.  We assume a delay--Doppler channel with maximum delay tap $l_{\tau}=20$ and Doppler tap $k_{\nu}=4$, which corresponds to maximum Doppler speed of $120$ Kmph. The  overhead  for pilot and guard symbols is approximately $1\%$ of an OTFS frame.
We observe that the BER reduces as ${\rm SNR}_p$ increases, providing more accurate channel estimation and better data detection. Moreover, the performance of OTFS with channel estimation is very close to the ideal case, when ${\rm SNR}_p=40$ dB (at least $20$dB higher than the data ${\rm SNR}_d$). Note that a large pilot power does not affect the peak transmit power as OTFS spreads each delay--Doppler symbol in the entire time--frequency plane thanks to the ISFFT operation. 

In Fig. \ref{sim4}, we perform comparisons of BER versus ${\rm SNR}_d$ for different Doppler frequencies with ${\rm SNR}_p=40$ dB, $l_{\tau}=20$, $\mathcal{T}=3\sigma$, and $4$-QAM. Consider UE speeds of 30, 120, and 500 Kmph corresponding to maximum Doppler tap $k_{\nu} = 1,4$, and $16$, respectively. We observe that the proposed estimation scheme exhibits highly similar performance under different Doppler frequencies except a slight performance improvement under higher Doppler frequencies (i.e., $k_{\nu} = 16$). This is due to the fact that more guard symbols and less data symbols are transmitted, leading to better data detection capability at higher ${\rm SNR}_d$. Since OTFS performs similarly at different frequencies, in the following, we consider only the UE speed of $120$ kmph. 
    
We next investigate the effect of the channel estimation threshold $\mathcal{T}$ on the system performance. Fix ${\rm SNR}_p=40$ dB. Fig. \ref{sim2} displays BER versus ${\rm SNR}_d$  with different $\mathcal{T}$.  We observe that the BER performance improves as $\mathcal{T}$ increases. 
%This behavior can be explained by the false alarm probability of the proposed channel estimator, which is equal to ${\rm{P}}_{{\rm{FA}}} = \exp(-\mathcal{T}^2/\sigma^2)$. 
For small threshold values, the path false detection probability is higher (i.e., it is more likely to detect non-existent paths), which degrades the BER performance. However, at the same time, increasing the threshold beyond a certain value may cause the likely miss detection of paths with small path-gains, resulting in performance loss. Hence, there is an optimal threshold to balance the false detection and miss detection probabilities. For the given system parameters, we observe that the optimal threshold is approximately $3\sigma$.      
\begin{figure}
\centering
\includegraphics[width=2.8in,height=2in]{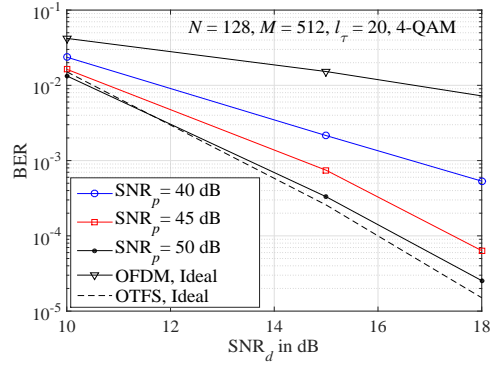}
\caption{BER versus ${\rm SNR}_d$: Fractional Doppler with full guard symbols.}
\label{sim3}
\end{figure}

\begin{figure}
\centering
\includegraphics[width=2.8in,height=2in]{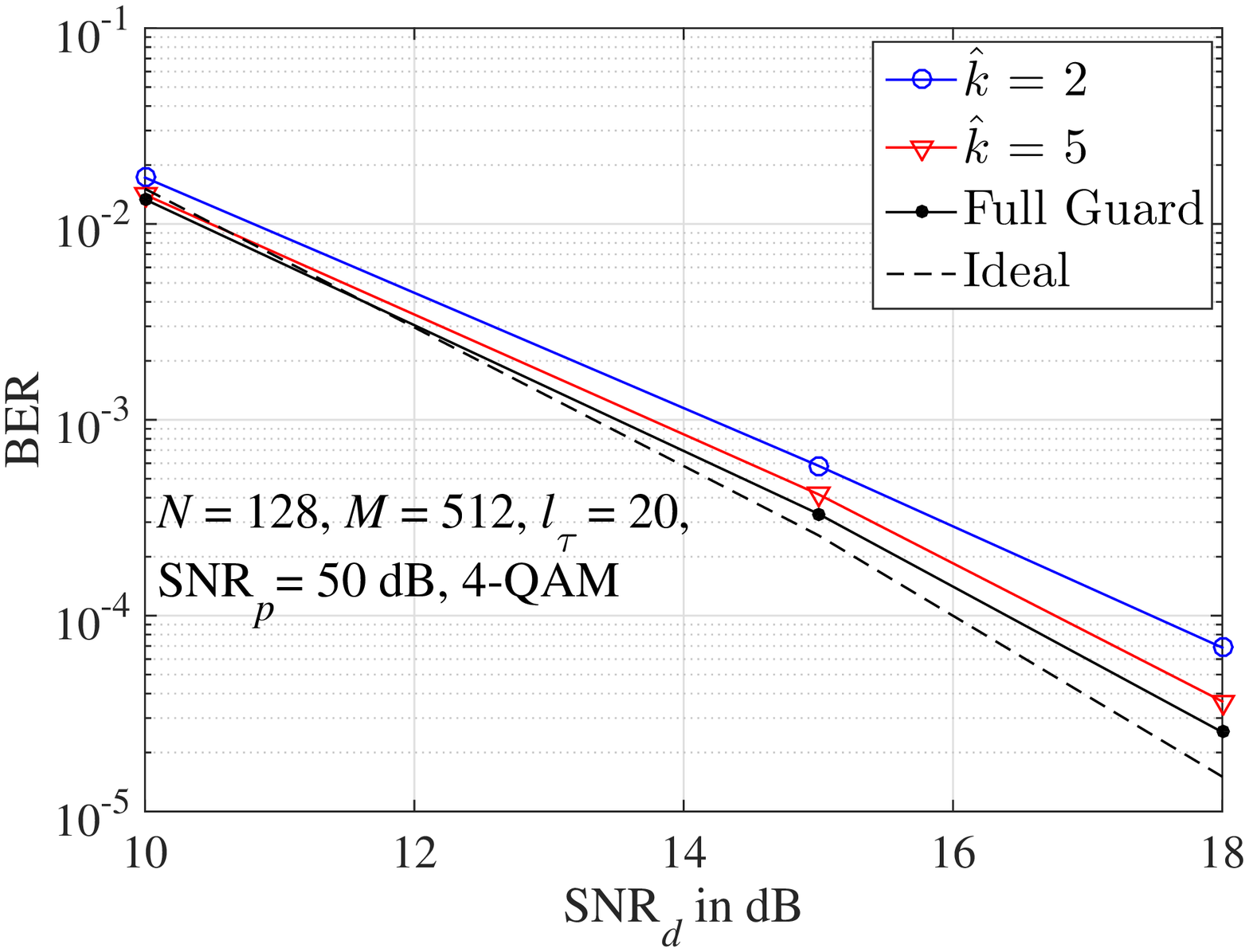}
\caption{BER versus ${\rm SNR}_d$: Fractional Doppler with reduced guard symbols.}
\label{sim5}
\end{figure}

\begin{figure}
\centering
\includegraphics[width=2.8in,height=2in]{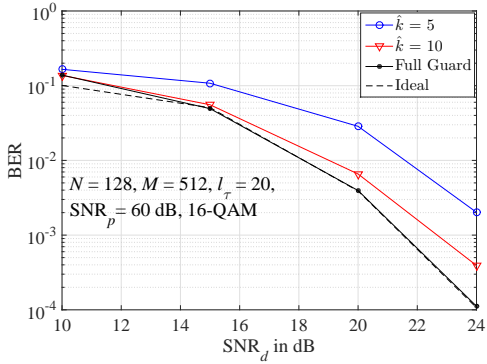}
\caption{BER versus ${\rm SNR}_d$: Fractional Doppler with reduced guard symbols for 16-QAM.}
\label{sim6}
\end{figure}

\begin{figure}
\centering
\includegraphics[width=2.8in,height=2in]{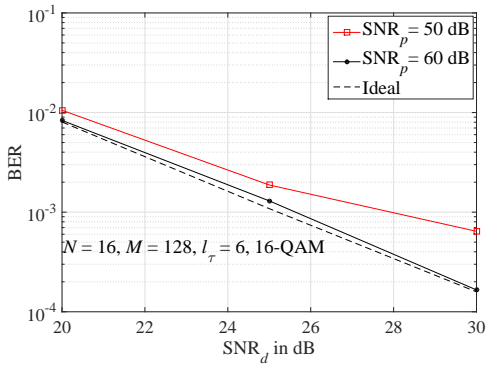}
\caption{BER versus ${\rm SNR}_d$: low latency communication}
\label{sim7}
\end{figure}

\subsection{The fractional Doppler case}
%We now consider channel estimation for the fractional Doppler case. 

Fig. \ref{sim3} shows the BER for different ${\rm SNR}_p$ with a threshold of $\mathcal{T}=3\sigma$. In this case, the pilot and guard symbols occupy approximately 8\% of an OTFS frame. Similar to the integer Doppler case, as more pilot power is used, the error performance is improved. As ${\rm SNR}_p=50$ dB, OTFS with our proposed embedded channel estimation attains similar performance as OTFS with known channel information. We can see that larger pilot power is required for channels with fractional Doppler shifts than integer Doppler shifts.
Last, we compare the BERs of OTFS with channel estimation and OFDM with known channel information and find that OTFS significantly outperforms OFDM, demonstrating the effectiveness of OTFS over delay--Doppler channels. 

In Fig. \ref{sim5}, we compare the BER performance of OTFS using the proposed channel estimation scheme with reduced guard symbols for $\hat{k}=2$ and $5$. Fix ${\rm SNR}_p=50$ dB, $\mathcal{T}=3\sigma$, and $4$-QAM. With $\hat{k}=2,$ and $5$, the overheads for pilot and guard symbols  are  roughly 1.5\% and 2.3\%, respectively, which are much less than the full guard symbols case (roughly 8\%). We observe that, as $\hat k$ becomes larger, the performance improves. In particular, with $\hat{k}=5$, the performance is very close to that with full guard symbols. For larger $\hat k$, smaller interference from neighboring data symbols improves the channel estimation accuracy. Hence, there is a tradeoff between spectral efficiency and error performance.

In Fig. \ref{sim6}, we illustrate the effectiveness of the proposed channel estimation schemes with full and reduced guard symbols, respectively, using $16$-QAM, ${\rm SNR}_p=60$ dB, and $\mathcal{T}=3\sigma$.  We see that with the higher pilot power (i.e., $60$ dB), the performance of our channel estimation scheme with full guard symbols is the same as that of the ideal case. Moreover, with 16-QAM, more guard symbols are required (i.e., $\hat{k}=10$, about 3.6\% guard symbols overhead) to achieve a performance close to the full guard symbols case, when compared to the 4-QAM case that adopts $\hat{k}=5$, about 2.3\% guard symbols overhead. This is due to the fact that the data detection of $16$-QAM case is more sensitive to the channel estimation and hence requires more guard symbols.

\begin{figure*}
\centering
\subfloat[Antenna 1 ($\times$: antenna $1$ data symbol)]{%
       \includegraphics[scale=0.7]{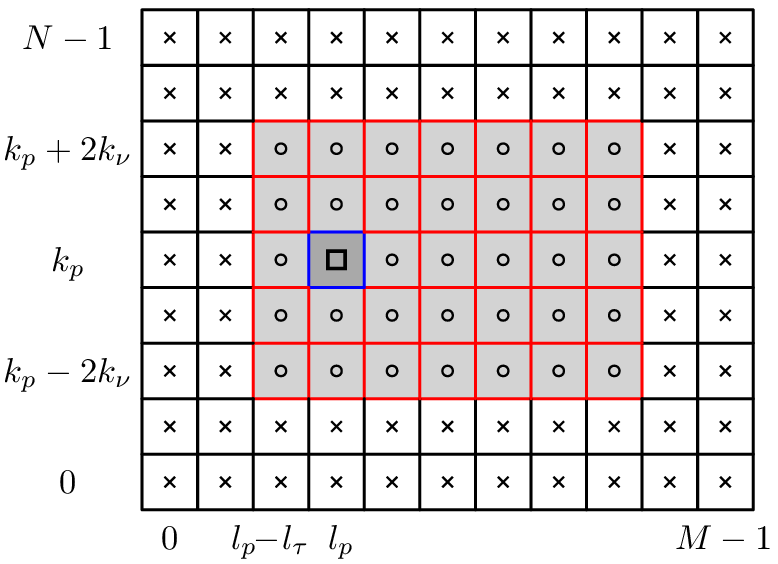}\label{mimo_grid_1}}
    \hspace{7mm}
  \subfloat[Antenna 2 ($\Diamond$: antenna $2$ data symbol)]{%
        \includegraphics[scale=0.7]{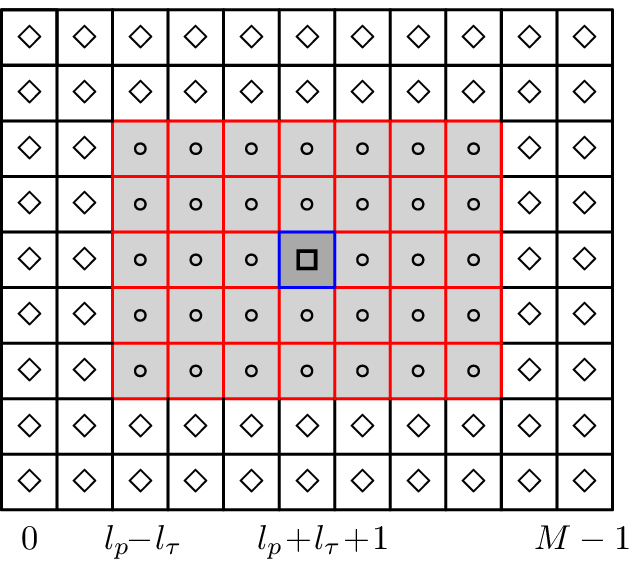}\label{mimo_grid_2}}
        \hspace{7mm}
        \subfloat[Antenna 3 ($\oplus$: antenna $3$ data symbol)]{%
        \includegraphics[scale=0.7]{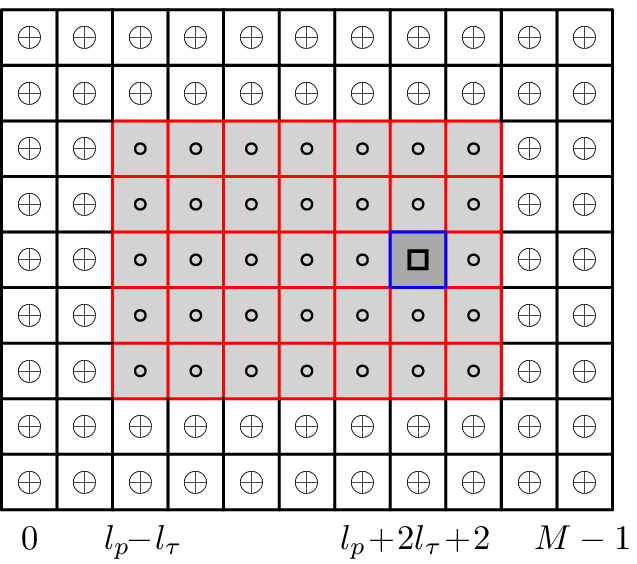}\label{mimo_grid_3}}
\caption{Tx pilot, guard, and data symbols for MIMO OTFS system ($\square$: pilot; $\circ$: guard symbols)} \label{mimo_grid}
\vspace{-2mm}
\end{figure*}

\begin{figure}
\includegraphics[scale=0.9]{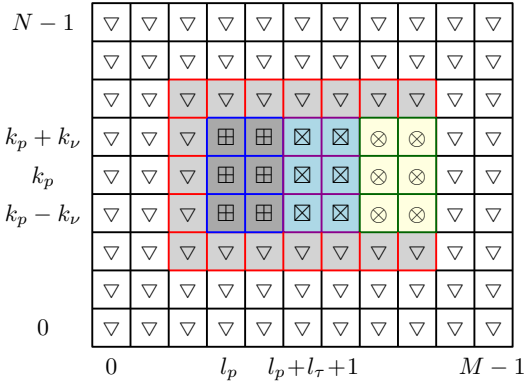}
\caption{Rx symbol pattern at one antenna of MIMO OTFS system ($\triangledown$: data detection, $\boxplus,\boxtimes,\otimes$: channel estimation for Tx antenna 1, 2, and 3, respectively)}
\label{mimo_rx_grid}
\end{figure}

\begin{figure*}
\centering
\subfloat[User 1 ($\times$: user $1$ data symbol)]{%
       \includegraphics[scale=0.7]{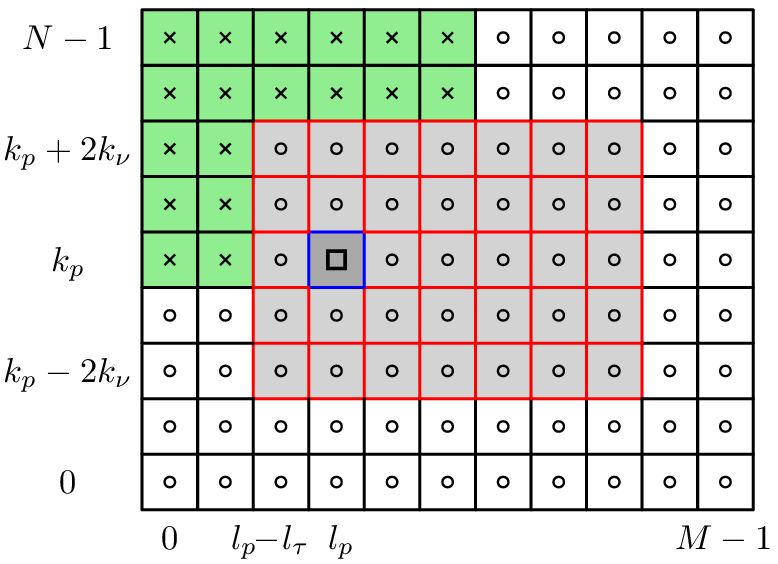}\label{muul_grid_1}}
    \hspace{7mm}
\subfloat[User 2 ($\Diamond$: user $2$ data symbol)]{%
        \includegraphics[scale=0.7]{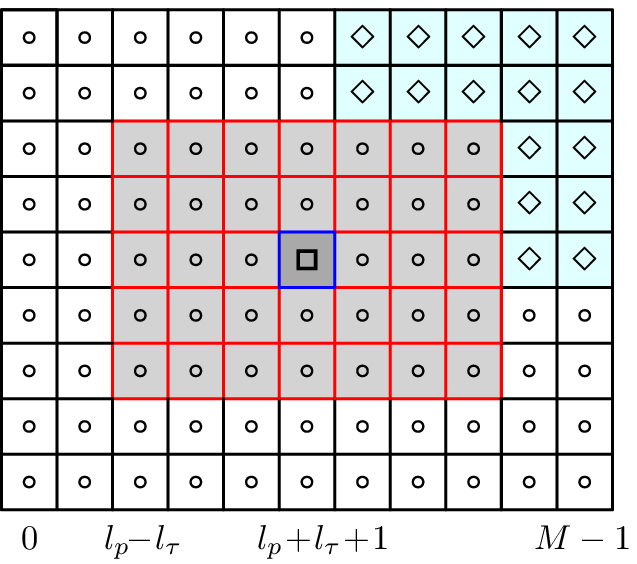}\label{muul_grid_2}}
        \hspace{7mm}
\subfloat[User 3 ($\oplus$: user $3$ data symbol)]{%
        \includegraphics[scale=0.7]{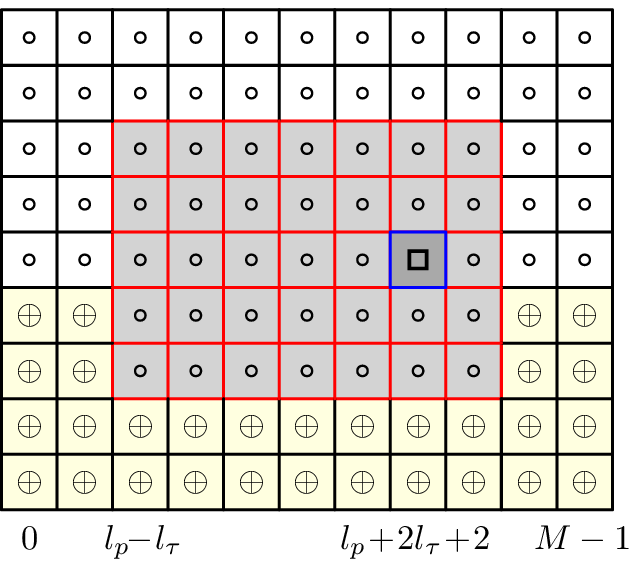}\label{muul_grid_3}}
\caption{Tx pilot, guard, and data symbols for multiuser uplink OTFS system ($\square$: pilot; $\circ$: guard symbols)} \label{muul_grid}
\vspace{-2mm}
\end{figure*}

\begin{figure}
\centering
\includegraphics[scale=0.9]{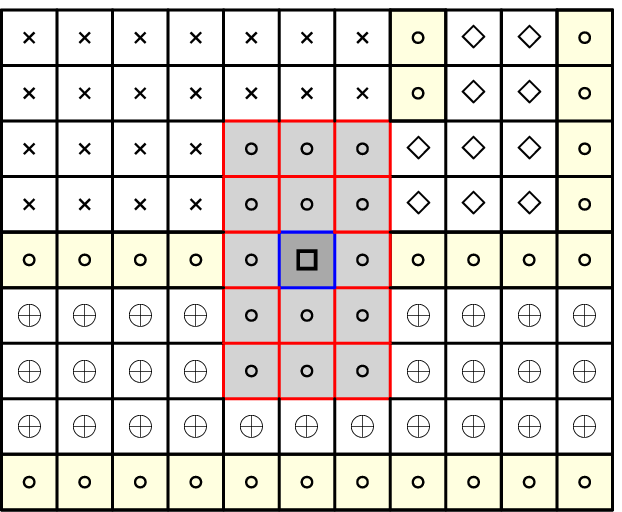}
\caption{Tx pilot and data arrangement for multiuser downlink OTFS system ($\square$: pilot; $\circ$: guard symbols; $\times, \Diamond,\oplus$: data symbols for users 1, 2, and 3, respectively)}
\label{mudl_pilot}
\end{figure}

\subsection{OTFS under low latency communications}
As next-generation wireless communications mostly require low latency communications, we next simulate the proposed OTFS channel estimation schemes under such scenario. 
Fig. \ref{sim7} shows the OTFS performance for low latency application with $N=16$ and $M=128$, corresponding to frame duration of $1.1$ ms. We consider the channel estimation scheme with full guard symbols as the reduced guard symbols case will not improve significantly the spectral efficiency with small $N$. We observe that the OTFS performance with channel estimation is very close to the ideal case with ${\rm SNR}_p=60$ dB. Hence, we can  conclude that the proposed channel estimation schemes are very efficient under low latency communications. 

\section{Extensions to MIMO and Multiuser Uplink/Downlink}
In this section, we extend our embedded channel estimation for point-to-point SISO OTFS systems to MIMO and multi-user uplink/downlink, respectively.  

%%%%%%%%%
\subsection{Point-to-point MIMO}
%%%%%%%%%%

In a MIMO system, each transmit (Tx) antenna arranges its own pilot, guard, and information symbols on the delay--Doppler grid for transmission (see Fig. \ref{mimo_grid}). The pilot symbol  is used to estimate the channels from that Tx antenna to each receive (Rx) antenna.  At each Rx antenna, different groups of received symbols are used for channel estimation from that Rx antenna to the Tx antennas, and for data detection from the Tx antennas. Moreover, the received symbols for data detection of the Rx antennas are jointly decoded using MP algorithm.  The symbol arrangements from the Tx antennas have to be carefully designed to facilitate the channel estimation and data detection at the Rx antennas. In the following, we describe one such arrangement.  

Consider a MIMO system with arbitrary $N_t \ge 1$ and $N_r \ge 1$. For ease of presentation, we consider channels with integer Doppler shifts and the case of fractional Doppler shifts is a straightforward extension. Inspired by our previous study in  Section III, we propose the following symbol arrangement $x^{n_t}[k,l]$ for the $n_t$-th Tx antenna ($n_t=1,\hdots,N_t$)     
\beqa 
x^{n_t}[k,l] = 
\begin{cases}
x_p &  ~~k=k_p, l =l_p+(n_t-1)(l_{\tau}+1),\\
0 & ~~k_p-2k_{\nu} \le k \le k_p+2k_{\nu},  \\ 
&~~\quad l_p-l_{\tau} \le l \le l_p+N_tl_{\tau} + N_t-1,  \\
x_d^{n_t}[k,l] &  ~~\text{otherwise} \nonumber 
\end{cases} \eeqa 
where $x_d^{n_t}[k,l]$ denotes the data symbol at location $[k,l]$ of $n_t$-th Tx antenna. We can see that the pilot symbols of the Tx antennas are sufficiently separated (by the maximum delay tap $l_{\tau}$ along the delay axis) so that they do not interfere with each other at the Rx antennas, as demonstrated in Fig. \ref{mimo_grid} for an exemplary MIMO system with three Tx antennas. 

At the $n_r$-th Rx antenna ($n_r=1,\hdots,N_r$), the received symbols $y^{n_r}[k,l], k_p-k_{\nu} \le k \le k_p+k_{\nu}, l_p+(n_t-1)(l_{\tau}+1) \le l \le  l_p+n_tl_{\tau} + n_t-1$, are used for channel estimation to the $n_t$-th Tx antenna. These received symbols are affected by the pilot signal of the $n_t$-th Tx antenna and by the channel between the $n_t$-th Tx and $n_r$-th Rx antennas only, as shown in Fig. \ref{mimo_rx_grid}. Hence, the channel estimation technique in Section III can be applied straightforwardly. The remaining received symbols of the $n_r$-th Rx antenna are functions of the data symbols from all the Tx antennas and thus a joint detection in \cite{Chock1} can be applied.  We omit the details for brevity. 

\subsection{Multiuser}

Consider a multiuser system, where single-antenna users communicate with base station in uplink or downlnk. The base station has either single or multiple antennas. In the following, we present embedded channel estimation schemes using Tx symbol arrangement for the users and base station.  
\subsubsection{Uplink} Consider single-antenna base station. We assume orthogonal resource allocation among the users. 

One example of the Tx symbol arrangements for three-user case is shown in Fig. \ref{muul_grid}. For each user, in each OTFS frame, the grid locations $[k,l], k_p-2k_{\nu} \le k \le k_p+2k_{\nu}, l_p-l_{\tau} \le l \le l_p+N_ul_{\tau} + N_u-1$ are used for pilot and guard symbols, where $N_u$ is the number of users. The pilot symbols of the users are located sufficiently apart at suitable locations as in the MIMO case. Moreover, each user occupies only a non-overlapping portion of the rest of the grid locations for its data transmissions with the remaining grid locations being used for zero symbols since orthogonal resource allocations is required, as shown in Fig. \ref{muul_grid}, where green, blue, and yellow grids contains data for Users $1$, $2$, and $3$, respectively. The data portion for each user depends on the resource requirement/allocation. Based on the Tx symbol arrangements, the base station exploits suitable received symbols for channel estimation and data detection for the users.  
\begin{remark}
When the base station has multiple antennas, the grid locations for pilot and guard symbols for the users remain intact. However, each user can exploit a larger portion, even full remaining grids for data transmissions, similar to the MIMO case.
\end{remark}

\subsubsection{Downlink}
Consider single-antenna base station, transmitting a pilot symbol being enclosed with guard symbols, similar to the point-to-point SISO case. This pilot signal is used by all the users to estimate the channel from itself to the base station. The rest of delay--Doppler grid locations is used for data transmissions to the users. Since orthogonal resource allocation is required, data symbols for users should be sufficiently separated using guard symbols to avoid inter-user interferences, as shown in Fig. \ref{mudl_pilot}, where yellow grids represent the guard symbols between users. Each user exploits appropriate groups of received symbols for channel estimation and detection of its own data. %The rest of the received symbols are neglected.  
% \begin{remark}
% When the base station has multiple antennas, we consider beamforming. To enable efficient beamforming, there are issues to be considered:
% \begin{itemize}
% \item[-] Channel state information (CSI) is required at the base station in order to transmit parallel beams to the users. CSI can be obtained by first transmitting the pilot only frame before the actual data and feedback the information from the users. One attractive approach that eliminates the extra overhead could be to exploit the uplink-downlink duality that is well studied in the context of massive MIMO systems ($\bf {CITE}$).
% \item[-] After the acquisition of CSI, the beamforming vectors in OTFS need to be designed in the delay-Doppler plane, where the channel relations are convolution nature, which is significantly different from OFDM with the element-wise product channel nature.
% \end{itemize}
% %In summary, beamforming for multiuser downlink OTFS is a very interesting direction for our future work.
% \end{remark}

\begin{table}[t]
\renewcommand{\arraystretch}{1}
\centering
\caption{Total number of pilot and guard symbols required for different embedded channel estimation schemes}
  \label{tab1}
  \begin{tabular}[c]{ | M{3.3cm} | M{3.9cm} | }
    \hline
    Method & Pilot + guard symbols\\ \hline
    SISO - integer Doppler & $(2l_{\tau}+1)(4k_{\nu}+1)$\\ \hline
    SISO - fractional Doppler full guard symbols & $(2l_{\tau}+1)(N)$\\ \hline
    SISO - fractional Doppler reduced guard symbols & $(2l_{\tau}+1)\left(4(k_{\nu}+\hat{k})+1\right)$\\ \hline
    MIMO - $N_t$ transmit antennas & $\left((N_t+1)l_{\tau}+N_t\right)\left(4(k_{\nu}+\hat{k})+1\right)$ \\ \hline
    Multiuser uplink  -  $N_u$ users with $1$ antenna & $\left((N_u+1)l_{\tau}+N_u\right)\left(4(k_{\nu}+\hat{k})+1\right)$ \\ \hline
    Multiuser downlink  -  base station with $1$ antenna & $\left(2l_{\tau}+1\right)\left(4(k_{\nu}+\hat{k})+1\right)$ \\ \hline
  \end{tabular}
  
\end{table}

Table \ref{tab1} summarizes the total number of pilot and guard symbols required for the different channel estimation methods in our paper.

%different threshold to account for the extra coefficient in the effective path gain. 
\section{Conclusion}
In this work, we have developed embedded pilot-aided OTFS channel estimation schemes. In particular, we arrange pilot, guard, and information symbols in the delay--Doppler grids to suitably avoid interference between pilot and data symbols. We design such arrangements for OTFS with ideal and rectangular pulses over channels with integer or fractional Doppler paths, respectively. 
At the receiver, channel estimation is performed based on a threshold method and the estimated channel information is used for data detection via a MP algorithm. We compare by simulations the error performance of OTFS using the proposed channel estimation schemes and OTFS with perfectly known channel information and observe only a marginal performance loss. Further, we show that OTFS with our channel estimation significantly outperforms OFDM with ideal channel information. Extensions of the proposed schemes to MIMO and multi-user uplink/downlink  have been presented. 

\section*{Acknowledgement}
This research work is supported by the Australian Research Council under Discovery Project ARC DP160101077. Simulations were undertaken with the assistance of resources and services from the National Computational Infrastructure (NCI), which is supported by the Australian Government.

\end{document}